%% file: main.tex
\documentclass[journal=jpccck,manuscript=article]{achemso}

\usepackage[version=3]{mhchem} % Formula subscripts using \ce{}
\usepackage[T1]{fontenc}       % Use modern font encodings
\usepackage[utf8]{inputenc}    % enables Å Ä Ö å ä ö to be displayed correctly
\usepackage{amsmath}
\usepackage{graphicx}
\usepackage{color}
\usepackage{verbatim}
\newcommand{\onlinecite}[1]{\nocite{#1}\citenum{#1}} 

\author{M.~R.~Mahani}
\affiliation{Department of Applied Physics, School of Engineering Sciences, KTH Royal Institute of Technology, Electrum 229, SE-16440 Kista, Sweden}
\email{mrmahani@kth.se}
\author{A.~Mirsakiyeva}
\affiliation{Department of Applied Physics, School of Engineering Sciences, KTH Royal Institute of Technology, Electrum 229, SE-16440 Kista, Sweden}
\author{Anna Delin}
\email{annadel@kth.se}
\affiliation{Department of Applied Physics, School of Engineering Sciences, KTH Royal Institute of Technology, Electrum 229, SE-16440 Kista, Sweden}
\alsoaffiliation{Department of Physics and Astronomy, Materials Theory Division, Uppsala University, Box 516, SE-75120 Uppsala, Sweden}
\alsoaffiliation{Swedish e-Science Research Center (SeRC), KTH Royal Institute of Technology, SE-10044 Stockholm, Sweden}
\title{Breakdown of Polarons in Conducting Polymers at Device Field Strengths}

\keywords{poly(p-phenylene), SSH Hamiltonian, electric field, polaron, bipolaron}

\begin{document}
\begin{abstract}
Conducting polymers have become standard engineering materials, used in many electronic devices. Despite this, there is a lack of understanding of the microscopic origin of the conducting properties, especially at realistic device field strengths. We present simulations of doped poly(p-phenylene) (PPP) using a Su-Schrieffer-Heeger (SSH) tight-binding model, with the electric field included in the Hamiltonian through a time-dependent vector potential via Peierls substitution of the phase factor. We find that polarons typically break down within less than a picosecond after the field has been switched on, already for electric fields as low as around 1.6~mV/{\AA}. This is a field strength common in many flexible organic electronic devices. Our results challenge the relevance of the polaron as charge carrier in conducting polymers for a wide range of applications.
\end{abstract}
\input{Introduction}

\input{Theory}
\input{Results}
\input{Conclusions}
% 
\input{Acknowledgment}
\bibliography{ppp}

%\newpage
%
%
%\begin{figure*}
%\centering
%\includegraphics[scale=0.47]{TOC.pdf}
%\caption{Table of Contents Graphic}
%\label{fig:TOC}
%\end{figure*}
%

\end{document}

%% file: Introduction.tex
\section{Introduction}
Conducting polymers have emerged as important materials for a number of very diverse applications such as, e.g., fuel cells, field effect transistors, organic light-emitting diodes (OLED), solar cells, computer displays, thermoelectric films, and in microsurgical tools.~\cite{Gratzel2001,Forrest2004,Heeger1988,Gustafsson1992,Balint2014,Vasseur2016}.
Materials made from conducting polymers have numerous engineering-friendly properties -- they are often easily synthesised, air stable, solution processable, flexible, and environmentally friendly.
In several of the applications mentioned above, the charge transport properties are of central importance~\cite{pandey2015surface, chen2014solvent, adhikary2013enhanced}. Charge transport, and charge carrier mobility in particular, is one of the fundamental physical processes that determine device performance.
Considering how widely spread organic conductors are today in society -- many people now use touch screens every day -- the understanding of these charge carriers is surprisingly schematic, and existing models in use claim different physical origins of observed features.\cite{heimel2016optical}.
Improved insight into the underlying physical processes is vital for improving device performance. Due to the large electron-phonon coupling in these one-dimensional systems, charge is thought to be transported mainly in the form of polarons and (possibly) bipolarons, in which trapped charges self-localise together with an associated structural distortion. The dynamics of these charge carriers is complex and depends on, e.g.,  temperature, electric field, disorder, and system dimensionality~\cite{Ribeiro2015,Ribeiro2013,Silva2012,Johansson2004,zhang2002microscopic}.

Here, we investigate theoretically charge transport as a function of electric field strength in the conducting polymer poly(p-phenylene) (PPP), also known as poly(1,4-phenylene). PPP was discovered in 1979\cite{ivory1979highly} and is recognised as a useful high-performance polymer due to its thermal~\cite{kovacic1987dehydro} and chemical~\cite{ivory1979highly} stability and its electrical and optoelectronic properties~\cite{grem1992realization, grem1992blue, grem1993electroluminescence}. 
PPP has a very simple structure consisting of interconnected phenyl rings and therefore also serves as an archetypal example of a conducting polymer. PPP can also be viewed as an ultrathin graphene nanoribbon. From a theoretical perspective, a lot of work on fundamental aspects of charge transport in conducting polymers have been performed on polyacetylene~\cite{Ribeiro2013,Silva2012,Johansson2004,Johansson2001,Basko2002} (PA).

The present investigations employ a generalization of the Su-Schrieffer-Heeger (SSH) Hamiltonian~\cite{su1979solitons, su1981fractionally, PhysRevB.22.2099, PhysRevB.21.2388, su1980soliton},  with an electric field introduced through a time-dependent vector potential~\cite{ono1990motion}. Additional complexity and various interactions not included in this model are certainly present in real conjugated polymer systems. 
% temperature, dimensionality, disorder etc.
Though the size and morphology of polymer films in real devices is more complex than what is presented in this work (a single chain), a bottom-up approach is of special interest for understanding the underlying physical processes, determining the relation between the motion of the excitations and the electric field strength.

In devices made from conducting polymers, the strength of the applied electric field varies depending on the type of device. In thermoelectric applications, the field is of the order of $\mu$V/\AA~\cite{bubnova2012tuning}, whereas in optoelectronic devices, field strenghts of typically a few mV/{\AA} are common.\cite{Ribeiro2013, lin2003triplet, gross2000improving, barth2001electron, yamamoto2005nanoscale, tutis2003internal}.We have used these experimental values as a guideline in the present work. Specifically, we address field strengths in the range 1~$\mu$V/\AA -- 5~mV/\AA. Our results indicate that polarons break down already at very moderate field strengths, common in devices. Also the bipolarons show significant instability in the upper range of the addressed electric field strengths. However, we would like to emphasize that our conclusions are drawn from the linear isolated chain. A chain which is not perfectly aligned with the electric field will of course feel a different electric field which would change the break-down value of the field. As regards the isolated chain approximation, hopping between chains seems to further lower the stability of polarons~\cite{Johansson2001}.

%% file: Theory.tex
\section{Theoretical model and computational details}
The system addressed throughout this work is a PPP polymer consisting of 20 six-membered rings with periodic boundary conditions.  
To describe the effect of an electric field on the fundamental excitations in a doped polymer, we study the time evolution of the Hamiltonian
\begin{equation} 
\label{hamiltonian}
\begin{split}
H_{\rm SSH} & = H_{\rm el}+H_{\rm latt}, \\
H_{\rm el} & =-\sum_n(t_0-\alpha(u_{n+1}-u_n))(e^{i\gamma A}c_n^\dagger c_{n+1}+e^{-i\gamma A}c_{n+1}^\dagger c_n), \\
H_{\rm latt} & =\frac{1}{2} \sum_nK(u_{n+1}-u_n)^2+\frac{1}{2} \sum_nM\dot{u}_n^2,
\end{split}
\end{equation}
which is a generalization of the original SSH Hamiltonian to account for the presence of an electric field~\cite{ono1990motion}. 
In the above Hamiltonian, $H_{\rm el}$ is the electronic part and $H_{\rm latt}$ is the lattice part. $H_{\rm el}$ contains, in addition to hopping terms, the electron-phonon coupling terms. The parameter $t_0$ is the hopping integral between nearest-neighbor carbon sites, $\alpha$ is the electron-phonon coupling constant for the $\pi$ electrons, $u_n$ is the displacement of the \textit{n}-th CH group\cite{comment1} from its equilibrium position, $c_n^\dagger$ ($c_n$) is the creation (annihilation) operator for a $\pi$ electron on the \textit{n}-th site, $K$ is the elastic constant associated with the $\sigma$ bonds and $M$ is the mass of the CH group.

The electric field is introduced in terms of a time-dependent vector potential, $A$, via Peierls substitution of the phase factor~\cite{peierls1955quantum, ono1990motion}. The parameter $\gamma$ is defined as $\gamma=ea/\hbar c$ with $e$, $a$, $c$ being the absolute value of the electronic charge, the lattice constant and the speed of light, respectively. The electric field, then, is given by the time derivative of the vector potential, i.e. $E=-(1/c)\partial A / \partial t$.

We use periodic boundary conditions to avoid end-point effects. This also allows us to study the dynamics of the excitations for a longer period of time, without having to use an unnecessarily large system. The periodic boundary conditions implies that the position of the polaron (bipolaron), after it forms, can be anywhere along the chain, depending on the initial conditions. If one removes the periodic boundary conditions, the polaron (bipolaron) localised in the middle of the chain has the lowest ground state energy~\cite{johansson2004nonadiabatic}.

A time dependent electric field uniform in space can be described with a vector potential that is independent of the space coordinates (a scalar potential describing the same electric field will not have this property). Therefore, introducing the electric field through a vector potential in the form of a complex phase factor to the phase integral is compatible with periodic boundary conditions.
In our simulations, we have set the parameters appearing in the Hamiltonian to the following values: $t_0=2.5$\,eV, 
$K=21$\,eV\AA$^{-2}$,
$\alpha=4.1$\,eV\AA$^{-1}$, 
$a=1.22$\,{\AA} and 
$M=1349.14$\,eVfs$^2$\AA$^{-2}$. 
These values are the same as those commonly used for polyacetylene (PA)~\cite{ono1990motion}. Our choice is motivated by the fact that the backbone of PPP is essentially identical to that of PA. Additionally, this set of parameters not only reproduce the electronic properties of PPP in zero field (measured energy gaps in short PPP oligomers\cite{diaz1981,botelho2011unified}), but it also allows us to compare our results with those of PA.
The PA structure using the same tight-binding Model has the energy level and band gap according to Ref.[~\onlinecite{su1979solitons, PhysRevB.22.2099, rossi1991minimum}], while using the same set of parameters for PPP, as first order of approximation, would give a different electronic structure and band gap in our calculations.

In principle, the band gap can be fine-tuned by introducing additional parameters for the electron-phonon coupling in aromatic rings~\cite{botelho2011unified}. However, such fine-tuning can be expected to have a relatively small effect on the overall charge carrier dynamics and therefore we have not included them in the present work. 
More elaborate models -- with, e.g., the electron-electron interaction included together with the already mentioned additional parameters for the electron-phonon coupling in aromatic rings -- will be the subject of future work.

The static initial-state geometry of the charged polymer is determined with the electric field set to zero, allowing polarons and bipolarons to form. By minimizing the ground state total energy of the system using the Hellman-Feynman force theorem within the adiabatic approximation with zero electric field, one arrives at
\begin{equation}
u_n=\frac{1}{2}(u_{n+1}+u_{n-1})+\frac{2\alpha}{K}\sum_k'{(\psi_k^*(n)\psi_k(n+1)-\psi_k^*(n-1)\psi_k(n))},
\label{scld}
\end{equation}
where $\psi_k(n)$ is an eigenfunction at site $n$. The sum is over all occupied states. The static initial-state geometry is then computed self-consistently by starting with an initial guess for the displacements, and then repeatedly applying equation~(\ref{scld}) until convergence, using the constraint that the total length of the system is constant~\cite{ono1990motion}.

In order to study the dynamics of the geometry-optimised charged excitations, we switch on the electric field at $t=0$ and linearly raise it to its maximum during the first 50~fs of the simulation. With this gradual onset of the electric field, we avoid disturbances due to sudden switching~\cite{kuwabara1991motion}. 
In principle, switching on the electric field before optimizing the doped system could be used to address processes in which the electrons are, for example, hopping from another chain into the system. However, there are studies for polyacetylene showing that applying a field above a certain value before optimizing the structure leads to nonspecific patterns rather than forming polarons or bipolarons.\cite{rakhmanova1999polaron} Such transport processes are deemed out of scope in the present work.

The evolution of the system is computed by solving the time-dependent Schr\"odinger equation 
\begin{gather}
\nonumber
i\hbar\frac{\partial\psi(n,t)}{\partial t}=(-t_0+\alpha(u_{n+1}-u_n))e^{i\gamma A}\psi(n+1,t)+\\
(-t_0+\alpha(u_{n}-u_{n-1}))e^{-i\gamma A}\psi(n-1,t),
\label{tdsch}
\end{gather}
and the equation of motion for the lattice displacement
\begin{gather}
\nonumber
M\ddot{u}_n(t)=F_n(t)=-K(2u_n-u_{n+1}-u_{n-1})+\\
\alpha\sum_k'{e^{i\gamma A}(\psi_k^*(n)\psi_k(n+1)-\psi_k^*(n-1)\psi_k(n))}+H.c.
\label{cem}
\end{gather}
 simultaneously.
In the equation of motion, equation~(\ref{cem}), the forces $F_n(t)$
are derived from the total potential, i.e. the sum of the electronic potential and the lattice harmonic potential.\\
The coupled differential equations are solved numerically using the procedure of Ono and coworkers~\cite{ono1990motion}, briefly outlined below. The solutions of the time-dependent Schr\"odinger equation are
\begin{equation}
\psi_k(t)=Te^{-i/\hbar\int_0^t\hat{h}(t^\prime)dt^\prime}\psi_k(0),
\end{equation}
where $\hat{h}(t)$ 
is the single particle Hamiltonian and $T$ the time ordering operator. 
To compute this expression numerically, time needs to be discretised. We choose the time step $\Delta t$ to be 0.025\,fs, which is very small on the scale of the bare phonon frequency $\omega_Q=\sqrt{4K/M}$ of the system.
With this choice, the variation of the Hamiltonian within a time step can be assumed to be negligible, and the electronic wave function can be written 
\begin{equation}
\psi_k(t_{j+1})=e^{-i\hat{h}(t_j)\Delta t/\hbar}\psi_k(t_j). 
\end{equation}
By expanding the electronic wave function in terms of the eigenfunction ($\phi_l$) and eigenvalues ($\epsilon_l$) of the single-particle Hamiltonian $\hat{h}(t)$ at each time step, the wave function becomes 

\begin{equation}
\psi_k(n,t_{j+1})=\sum_l \bigg[\sum_p\phi_l^*(p) \psi_k(p,t_j)\bigg] e^{-i\epsilon_l\Delta t/\hbar}\phi_l(n). 
\end{equation}
This set of coupled equations can be numerically integrated using the following algorithm: 
\begin{equation}
u_n(t_{j+1})=u_n(t_{j})+\dot{u}_n(t_{j})\Delta t, \qquad
\dot{u}_n(t_{j+1})=\dot{u}_n(t_{j})+\frac{F_n({t_j})}{M}\Delta t\;,
\end{equation}
resulting in the pertinent time-dependent electronic wave functions.

To analyze the motion of the excitations, the polaron and bipolaron positions and velocities need to be computed. The position of the excitation is defined by considering the center of mass $x_c$ for the excess charge density $\rho_n$, taking the periodic boundary conditions into account.\cite{ono1990motion} Thus, 
\begin{equation}
  x_c=
  \begin{cases}
    N\theta/2\pi, & \left<\cos{\theta_n}\right> \geq 0  \quad \text{and} \quad \left<\sin{\theta_n}\right> \geq 0,\\
    N(\theta+\pi)/2\pi, & \left<\cos{\theta_n}\right> \leq 0,\\
    N(\theta+2\pi)/2\pi, & \text{otherwise},
  \end{cases}
\label{velo}
\end{equation}
where
\begin{gather}
 \nonumber
 \left<\cos{\theta_n}\right>=\sum_n{\rho_n \cos{(2\pi n/N)}},  \quad 
 \left<\sin{\theta_n}\right>=\sum_n{\rho_n \sin{(2\pi n/N)}},\\
 \nonumber
\theta=\arctan{\left( \frac{\left<\sin{\theta_n}\right>}{\left<\cos{\theta_n}\right>}\right)},
\end{gather}
and the excess charge density $\rho_n$ is given by
\begin{equation}
\rho_n(t)=\sum_k'{\mid \psi_k(n,t)\mid^2-1}.
\label{excha}
\end{equation}
From the computed $x_c$, the velocity is calculated as an average velocity over 400 time steps according to
\begin{equation}
v(t_j)=\frac{x_c(t_j)-x_c(t_j-400\Delta t)}{400\Delta t}.
\label{velocity}
\end{equation}

As already mentioned, our model does not include the effects of electron-electron correlations. These effects can be implemented through the on-site and off-site Hubbard terms, similar to what has been done for PA~\cite{ozawa2000dynamics,PhysRevB.24.2168, lin2006multiple}.
Alternatively, the density matrix renormalization group (DMRG) could be used for addressing the interactions~\cite{jeckelmann1998density}. 

%% file: Results.tex
\section{Results and discussion}
Our simulations reveal that the behavior of the charge carriers depend sensitively on the electric field strength. There is a weak field regime with well-localised polaronic sonic or supersonic states, and a strong field regime, where the charge and lattice degrees of freedom decouple. In Table~\ref{tab:critical_fields} we summarise our computed critical field strengths. 
\begin{table}[ht]
    \centering
    \begin{tabular}{c|c|c}
                               & PPP$^{+}$ & PPP$^{2+}$ \\ \hline
       sonic -- supersonic     & 0.15-0.155  & 0.6-0.65 \\
       supersonic -- breakdown & 1.4-1.6     & 3.2-3.3 \\
    \end{tabular}
    \caption{Summary of critical field strengths for polarons (PPP$^+$) and bipolarons (PPP$^{2+}$) in PPP. The electric field strengths are given in mV/$\mbox{\AA}$.}
    \label{tab:critical_fields}
\end{table}
Interestingly, we find that polarons break down already at around 1.6~mV/\AA. This is a field strength present in many flexible electronics devices. Below, we describe the polaron and bipolaron properties as a function of field strength in more detail. We begin with the static regime, i.e. when the electric field is zero. The static regime is the starting point in all our simulations. 
\begin{figure*}
\begin{minipage}{1.0\linewidth}
\centering
\includegraphics[scale=0.25]{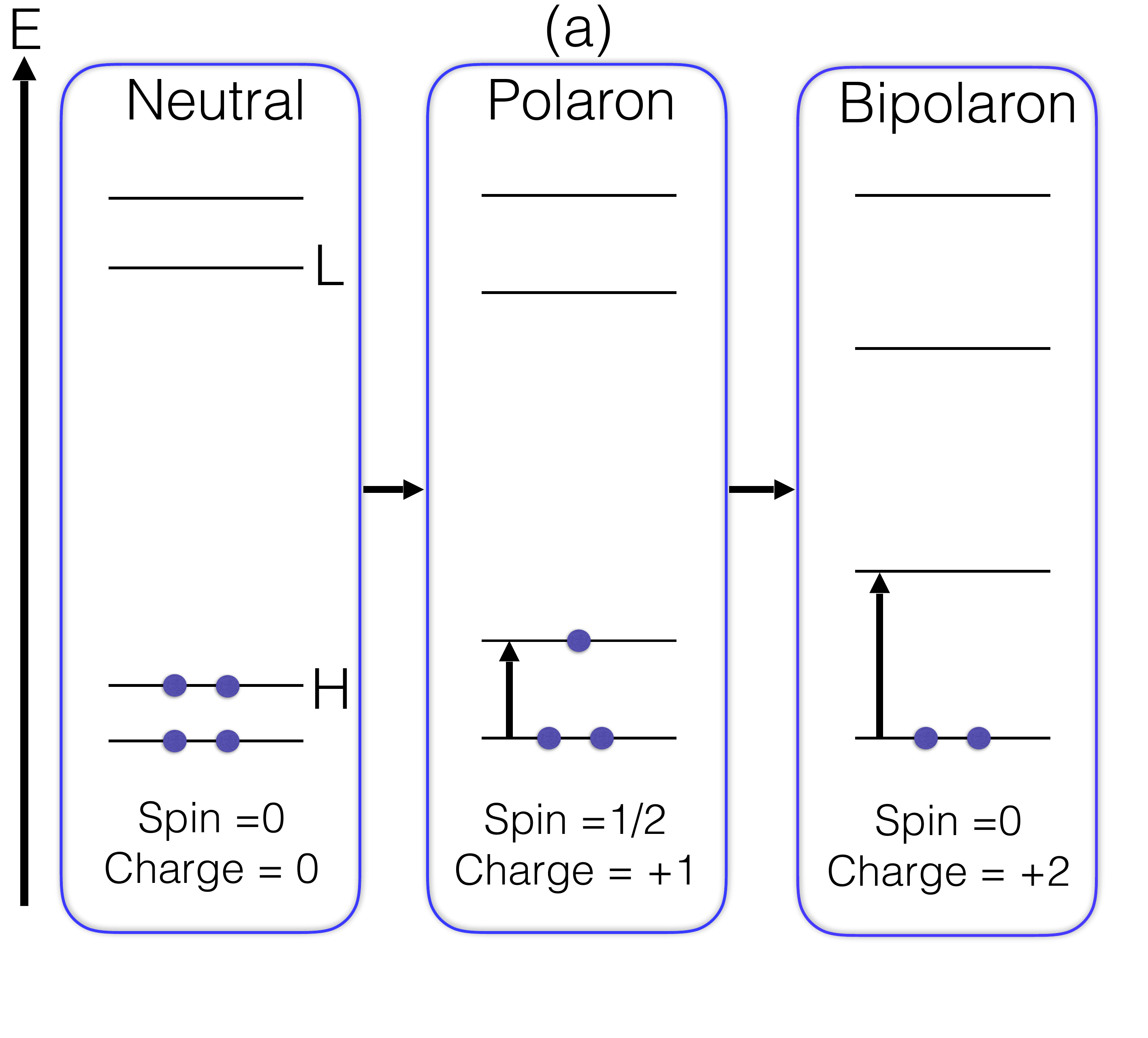}
\hspace{-3.50mm}
\includegraphics[scale=0.25]{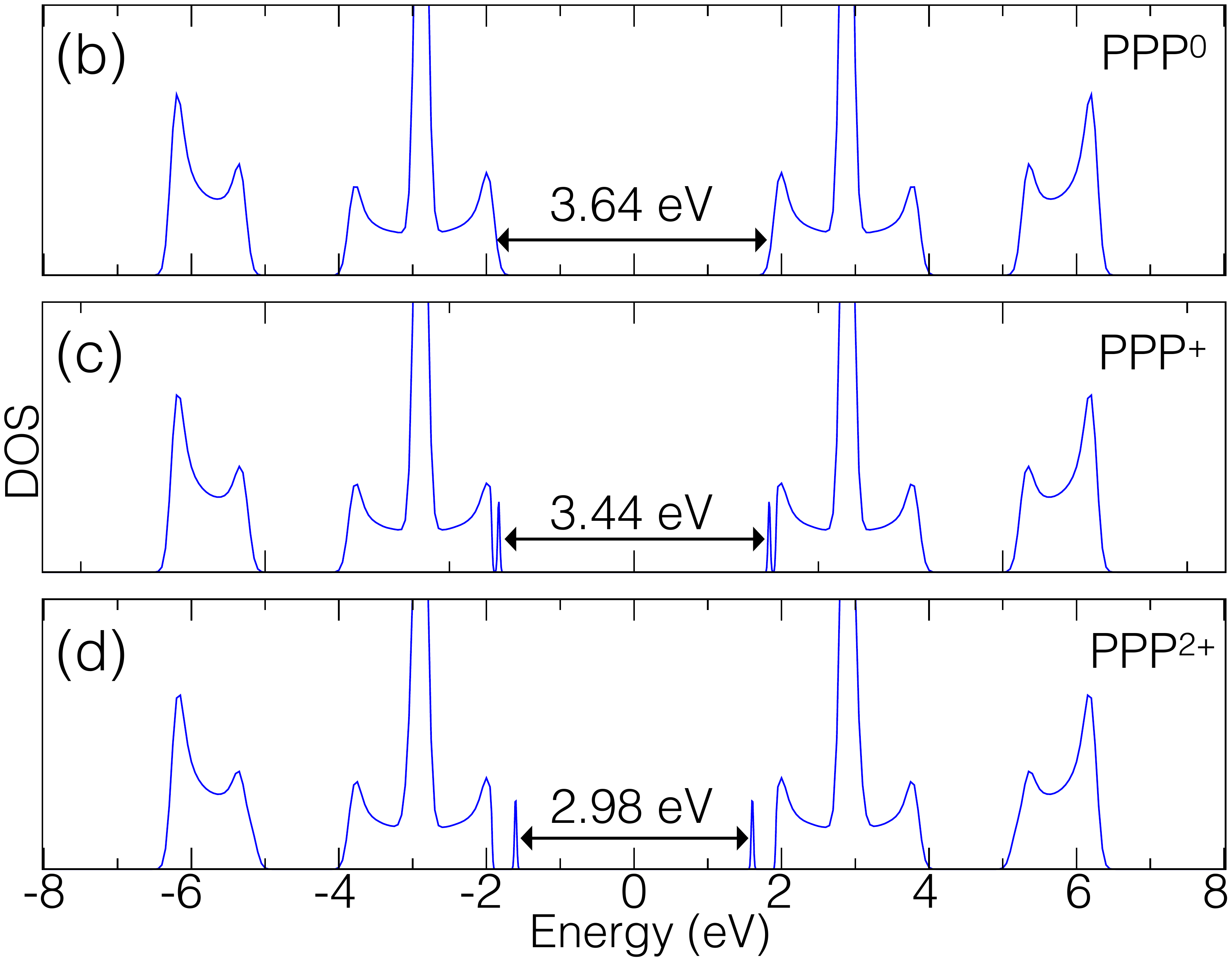}
\hspace{1.50mm}
\includegraphics[scale=0.47]{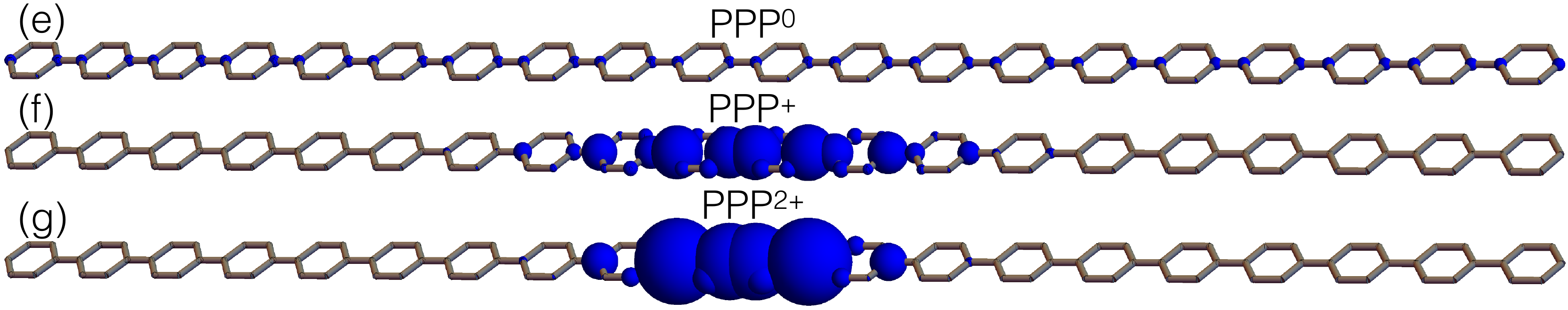}
\end{minipage}
		
\vspace{0.00mm}

\caption{(Colour online) Level structures, densities of states (DOS) and charge distributions of neutral and p-doped PPP. (a) The level structures in the gap for three cases. (b)--(d) DOS of neutral (PPP$^0$), singly charged (PPP$^+$) and doubly charged PPP (PPP$^{2+}$), respectively. (e)--(g) Visualization of the charge density corresponding to the highest occupied level (neutral case) and lowest in-gap (bi)polaronic levels, respectively. The radius of each blue sphere is proportional to the wave function amplitude on that site. The order parameter indicating the bond change coincides with the area of the polymer that the charge is localised to. The Fermi level for the neutral case is at the top of the valence band, where the DOS has finite value below the zero energy.}

\label{fig:pol.prop}
\end{figure*}

Schematic diagrams of the pertinent energy levels in the system are shown in Fig.~\ref{fig:pol.prop}a. The leftmost panel shows the neutral case (PPP$^0$). Here, each level is doubly occupied up to the Fermi energy. H stands for the highest occupied level, and L for the lowest unoccupied level. The middle panel shows the singly doped case (PPP$^+$), in which one electron has been removed. Two polaronic levels appear in the gap, of which the lowest one is singly occupied and the higher one unoccupied. Finally, the rightmost panel shows the bipolaronic case (PPP$^{2+}$). Here, two electrons have been removed. Compared to the polaronic case, the emerging levels appear deeper in the gap and they are both unoccupied. 

% dos
The calculated density of states (DOS) of PPP$^0$, PPP$^+$ and PPP$^{2+}$ polymers are shown in Fig.~\ref{fig:pol.prop}b-d, respectively. We see that the energy gaps and the shapes of the DOS for PPP$^0$ and PPP$^+$ are the same as in previous studies~\cite{botelho2011unified}. For the bipolaronic case, we find that the gap narrows to 2.98\,eV while it is 3.44\,eV for the singly charged case. Compared to the polaronic state, the bipolaron is more localised and deeper in the gap. 

In Fig.~\ref{fig:pol.prop}e-g, the charge densities of the three cases are visualised. For PPP$^0$, we show the charge density of the highest occupied level, whereas for PPP$^+$ and PPP$^{2+}$, we show the lowest in-gap level which represents the (bi)polaronic states. The radii of the blue spheres at each site correspond to the charge distribution of this level at that site. We see that the polaron extends over about six rings, whereas the bipolaron is about four rings wide. Our calculations thus confirm that the wave function of the bipolaron is more localised than that of the polaron, which agrees with the observed deeper gap states in the DOS (Fig.~\ref{fig:pol.prop}d).

It is relevant to compare our results with a recent DFT study~\cite{heimel2016optical} where a density functional with pure Hartree-Fock exchange in the long-range and pure Perdew-Burke-Ernzerhof (PBE) exchange in the short range was employed. In these calculations, the dihedral angles were also optimised, giving an interring rotation of about 39$^\circ$ in the neutral polymer, and an oscillating interring rotation ranging from around 22$^\circ$ to  50$^\circ$ inside the polaron. The long-range exchange tail allows a polaron to form. At the same time, shifts of the energy levels in and around the polaron due to the excess charge is accounted for. Due to these shifts, the gap states are no longer symmetrically positioned within the gap. The band gap in the neutral polymer in the DFT-based computation\cite{heimel2016optical} appears to be about 6\,eV, which is significantly larger than the band gap in our model (3.64\,eV), what experimental data for short isolated oligomers suggest\cite{diaz1981,botelho2011unified}, as well as what recent $GW$ computations find (3.95\,eV)\cite{puschnig2012band}.
The discrepancy between the typical experimental value of the gap in PPP (2.8\,eV) and the band gap in our model (3.64\,eV) warrants a few comments.
The value 2.8\,eV refers to the optical gap, and is determined from ultra-violet (UV) spectral measurements on a PPP film\cite{eckhardt1989electronic}, i.e., for PPP molecules not in vacuum. Interestingly, $GW$ calculations give that when a PPP chain is adsorbed on graphene at a distance of 4.0 \AA, the gap is renormalized from 3.95\,eV to 2.7\,eV\cite{puschnig2012band}. Thus, it appears that the environment heavily influences the value of the gap, and calculations for an isolated PPP chain cannot be compared directly with measurements on a PPP film, solution, or matrix.
The optical spectrum of a conjugated polymer typically also has a wide absorption band due to intrinsic disorder, which leads to significant uncertainty in the determination of the optical band gap. Furthermore, the electronic gap, which is the gap our model refers to, is very difficult to measure directly and is larger than the optical gap due to excitonic interaction. Finally, we also mention that in our model, the interring dihedral angles are assumed to be zero in all cases, which may somewhat affect the band gaps.

The visible changes in the bond lengths after doping the system in the mentioned DFT study~\cite{heimel2016optical} covers approximately six rings which compares well with our results for the polaron. In the DFT study, the bipolaron could not be stabilised, due to Coulomb repulsion. In general, including Coulomb repulsion in the SSH model is expected to weaken the bipolaronic state or cause it to become unstable, depending on the Coulomb interaction parameters.

\begin{figure*}
\begin{minipage}{1.0\linewidth}
\centering
\includegraphics[scale=0.32]{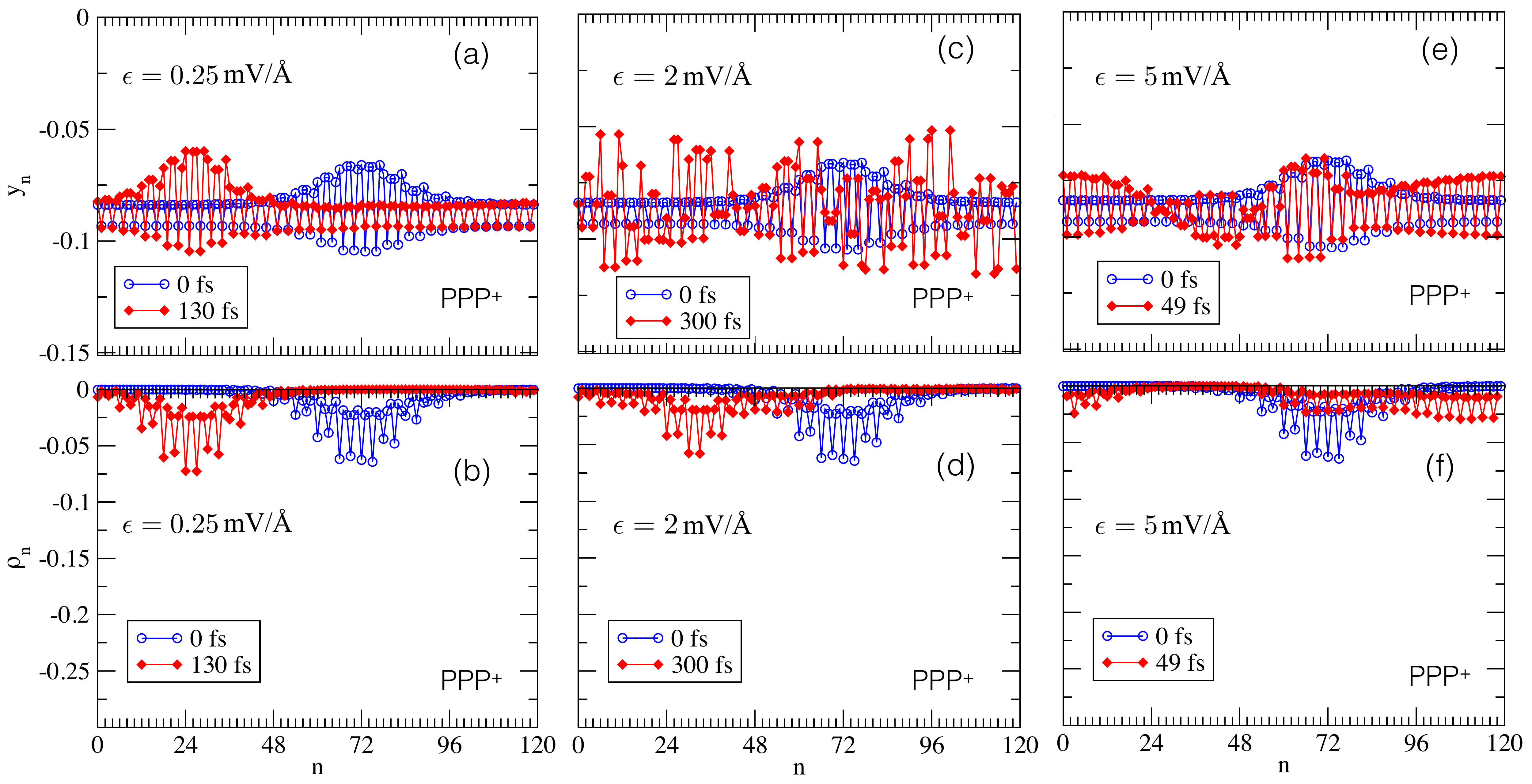}
\hspace{1.50mm}
\includegraphics[scale=0.48]{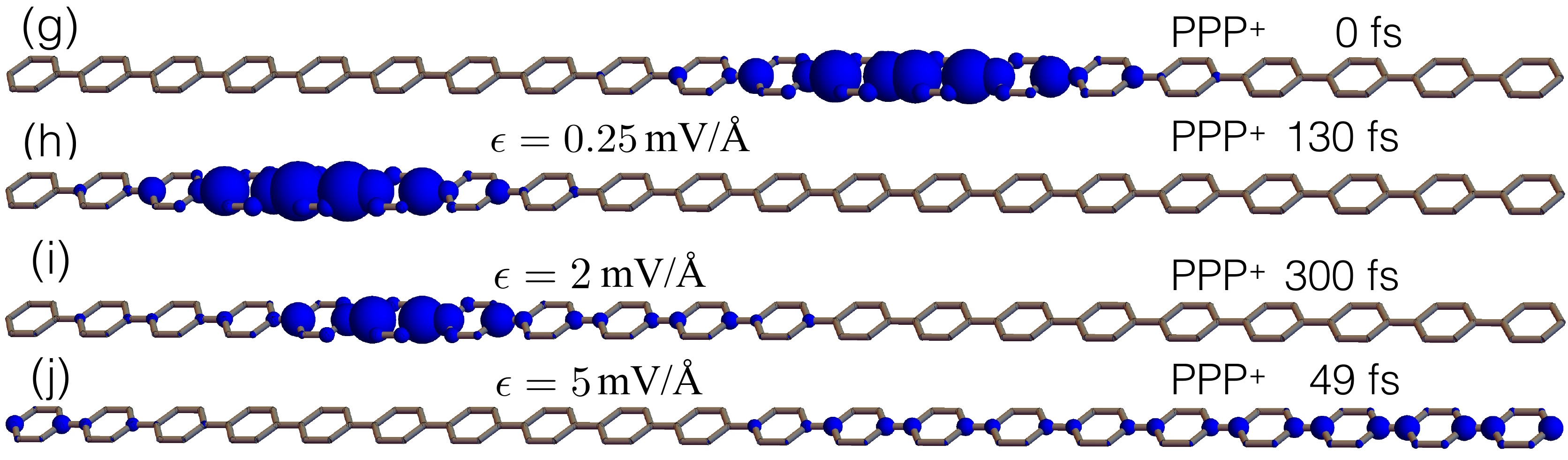}
\end{minipage}

\vspace{0.00mm}

\caption{(Colour online) The time evolution of the lattice distortion, excess charge, and polaronic-state wave function of p-doped PPP after applying 0.25, 2 and 5~mV/$\mbox{\AA}$ electric fields. (a) and (b): The evolution of the lattice distortion and the excess charge in 0.25~mV/$\mbox{\AA}$ electric field, respectively. (c), (d) and (e), (f): Similar to (a), (b) but in 2 and 5~mV/$\mbox{\AA}$ fields, respectively. The blue line with circle is for $t=0$ and the red line with filled-diamond is for $t$. (g) The polaron wave function at $t=0$ for all electric fields. The electric field is switched on at $t=0$ and gradually ramped up during the first 50~fs of the simulation. The size of the blue spheres in panels g-j is according to Fig.~\ref{fig:pol.prop}. (h)-(j) The evolution of the polaron wave function in 0.25, 2 and 5~mV/$\mbox{\AA}$ electric fields at different times.}
\label{fig:wfield}
\end{figure*}
We now turn to the real time dynamics of polarons as a function of electric field strength.
In Fig.~\ref{fig:wfield} we show the time evolution of the lattice distortions, excess charge and polaronic-state wave function of p-doped PPP after applying 0.25, 2 and 5~mV/$\mbox{\AA}$ electric fields. 
The weak field is chosen such that the charge and lattice distortions are coupled throughout the simulation time (up to 2~ps). The intermediate field is above the threshold for which the charge and the lattice distortions decouple. Finally, the strong field demonstrates how the polaron breaks down already at a very early stage in the simulation, just $t=49$~fs after initialisation. 

Figs.~\ref{fig:wfield}a and~\ref{fig:wfield}b show the time evolution of the displacements, $y_n=u_{n+1}-u_n$, and the excess charge $\rho_n(t)$, defined in equation~(\ref{excha}), for a polaron in a 0.25~mV/$\mbox{\AA}$ electric field. 
The blue circles (red filled diamonds) show a snapshot of the calculated quantity at $t=0$ ($t=130$~fs).  After 130~fs, the positively charged polaron has moved to the end of the chain in the direction of the applied field. Clearly, the displacement and the excess charge are coupled to one another and move as one entity. This remains the case for the entire simulation time, up to 2~ps.

Figs.~\ref{fig:wfield}e and~\ref{fig:wfield}f are similar to Figs.~\ref{fig:wfield}a and~\ref{fig:wfield}b but for an electric field of 5~mV/$\mbox{\AA}$. In this rather strong electric field the polaron charge cloud decouples from the lattice deformations.
As one can see from panels~\ref{fig:wfield}e and~\ref{fig:wfield}f, after only 49~fs the excess charge density is spread through the polymer and is thus completely detached from the lattice deformations. 
For electric fields strengths just above the stability threshold, here examplified iwth the field strengths 2~mV/$\mbox{\AA}$, the polaron breaks down in a similar way, but it takes longer time. This is illustrated in Figs.~\ref{fig:wfield}c-d. For long simulation times, we observe that polarons may occasionally reform and break down again.

Figs.~\ref{fig:wfield}g-j illustrate the corresponding eigenstates, showing the distribution of the charges on the atomic sites and their movements after applying the electric field. In our nearest-neighbor tight-binding model, the electric field forces the charge to hop between the nearest sites along the direction of the field. The size of the sphere around each site is proportional to the wave function amplitudes of that site at a given time. The excitations in panels~\ref{fig:wfield}h (weak field), continue to move as an entity along the chain even for very long simulation times. 
This picture does not hold for 2 and 5~mV/$\mbox{\AA}$ fields, where the polaron dissociates. This is clear from panels~\ref{fig:wfield}i, where the polaron after about 300~fs under 2~mV/$\mbox{\AA}$ electric field is delocalized. The localized charge spreads quickly (within only 49~fs) over a large part of the chain in 5~mV/$\mbox{\AA}$ electric field. Similar to the charge dynamics in polyacetylene (PA)~\cite{liu2007polaron}, we observe charge induced lattice deformation at later times after the initital polaron breakdown.

\begin{figure}
\centering
\includegraphics[scale=0.31]{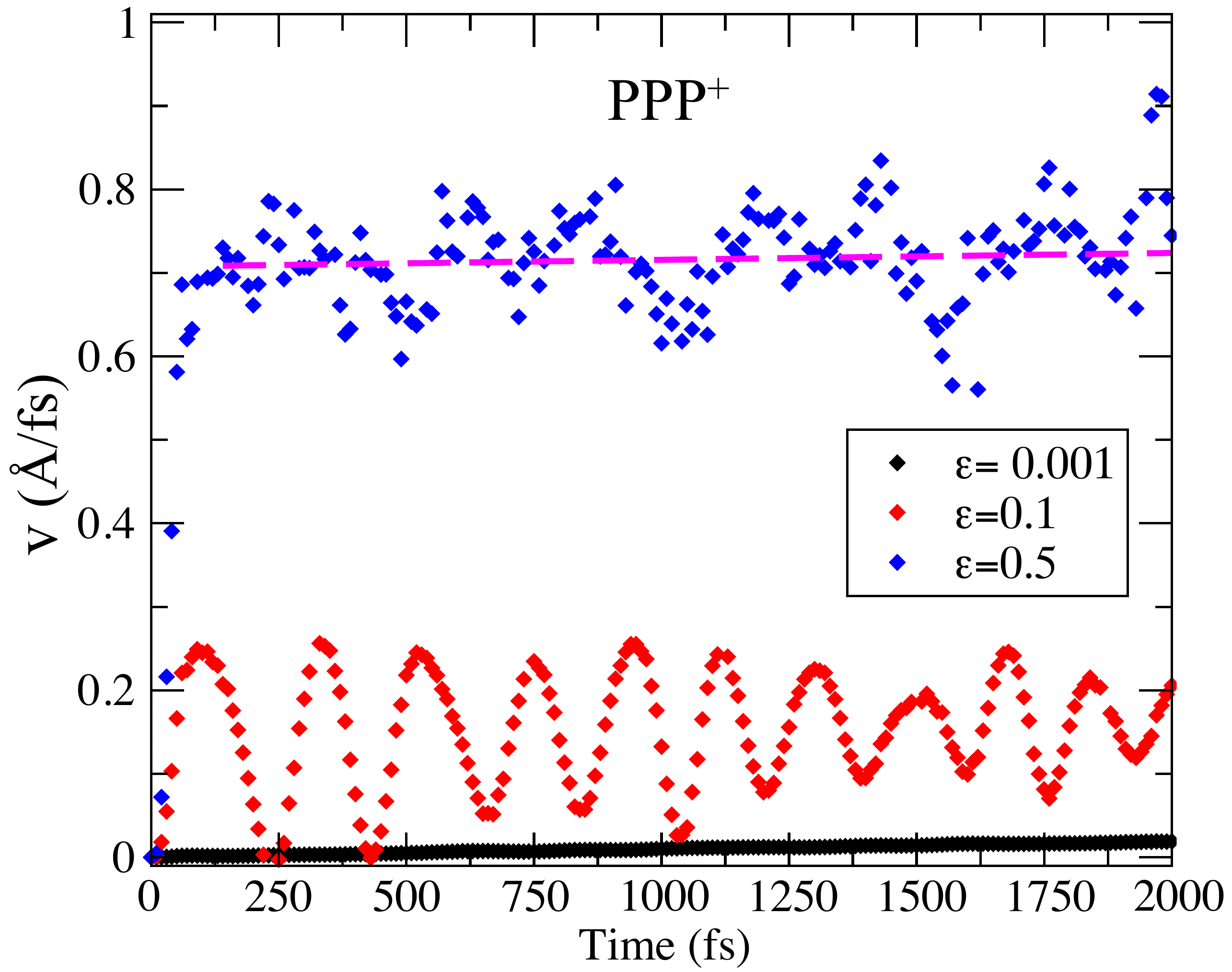}
\caption{(Colour online) The polaron velocity as a function of time for three representative electric field strengths $\epsilon$ in units of~mV/$\mbox{\AA}$. The magenta dashed line is a guide for the eye.}
\label{fig:vol}
\end{figure}

Fig.~\ref{fig:vol} exemplifies the polaron velocity at three different fields strengths: 
1~$\mu$V/\AA, 0.1~mV/\AA, and 0.5~mV/\AA. 
For all these field strengths the charge and the lattice are not decoupled and therefore the definition of the velocity (equation~(\ref{velocity})) holds. As one can deduce from Fig.~\ref{fig:vol}, the polaron velocity depends on the strength of the electric field in a highly nonlinear way. For the very weak field 1~$\mu$V/\AA, which is of the same order as the electric field in a thermoelectric device, the pumped energy via electric field does not move the polaron along the polymer chain, throughout our simulation (up to 2~ps).  When the field strength is increased but kept in the sonic regime, we observe significant oscillations in the velocity of the polaron, i.e. the polaron reaches a maximum speed and slows down -- in fact almost stops -- repeatedly. This regime is the result of coupling between charge and the acoustic phonons. For other fields in this regime, the velocities oscillate approximately around the same saturation velocity (not shown).

Finally, in the supersonic regime, where the charge instead couples to the optical phonons~\cite{johansson2004nonadiabatic}, the polaron velocity reaches a higher saturation value compared to sonic regime. In our simulations, the saturation velocity is attained after about 130~fs. After this short period of time, the energy pumped into the system by the electric field dissipates to the lattice vibrations at the same rate, so that the acceleration of the polaron becomes zero. The dissipation is continuous due to the classical description of the lattice in the SSH model. 

The saturation velocity in the supersonic regime is about three times larger (around 0.7~\AA/fs) than the maximum velocity in the sonic regime (around 0.25~\AA/fs). For other fields in the supersonic regime, the saturation velocity is similar to the one shown. The corresponding velocities for PA~\cite{johansson2004nonadiabatic} are about 0.45~\AA/fs in the supersonic regime and 0.13~\AA/fs in the sonic regime. Since the sound velocity, defined as $v_s=(a/2)\sqrt{4K/M}$ within this model, is similar in PA and PPP (due to the use of the same set of parameters) we believe the difference originates from the different geometries of these two polymers.
According to our simulations, the threshold electric field between the sonic and supersonic regimes lies between 0.15~mV/$\mbox{\AA}$ and 0.155~mV/$\mbox{\AA}$. This is slightly larger than what has been theoretically predicted for PA (0.135~mV/$\mbox{\AA}$--0.14~mV/$\mbox{\AA}$).\cite{johansson2004nonadiabatic}. 

In addition, we have also calculated the potential, electronic, kinetic and total energy differences, i.e. the energy at time $t$ subtracted by its counterpart at $t=0$, of the system. Our calculations show that the total energy of the system is increasing with time. This is expected, since the system is not connected to any external heat bath. In the weak field regime, the potential energy $(1/2)\sum_nK(u_{n+1}-u_n)^2$ is in phase but with opposite sign to the electronic energy $\sum_l'\sum_k'C_{k,l}\epsilon_l$. Therefore, the energy pumped into the systems via the electric field goes through the electrons to the lattice. In other words, in the weak field regime the potential energy increases and the electronic energy decreases, whereas in the strong field the electronic energy increases.
\begin{figure*}
\begin{minipage}[h]{1.0\linewidth}
\centering
\includegraphics[scale=0.4]{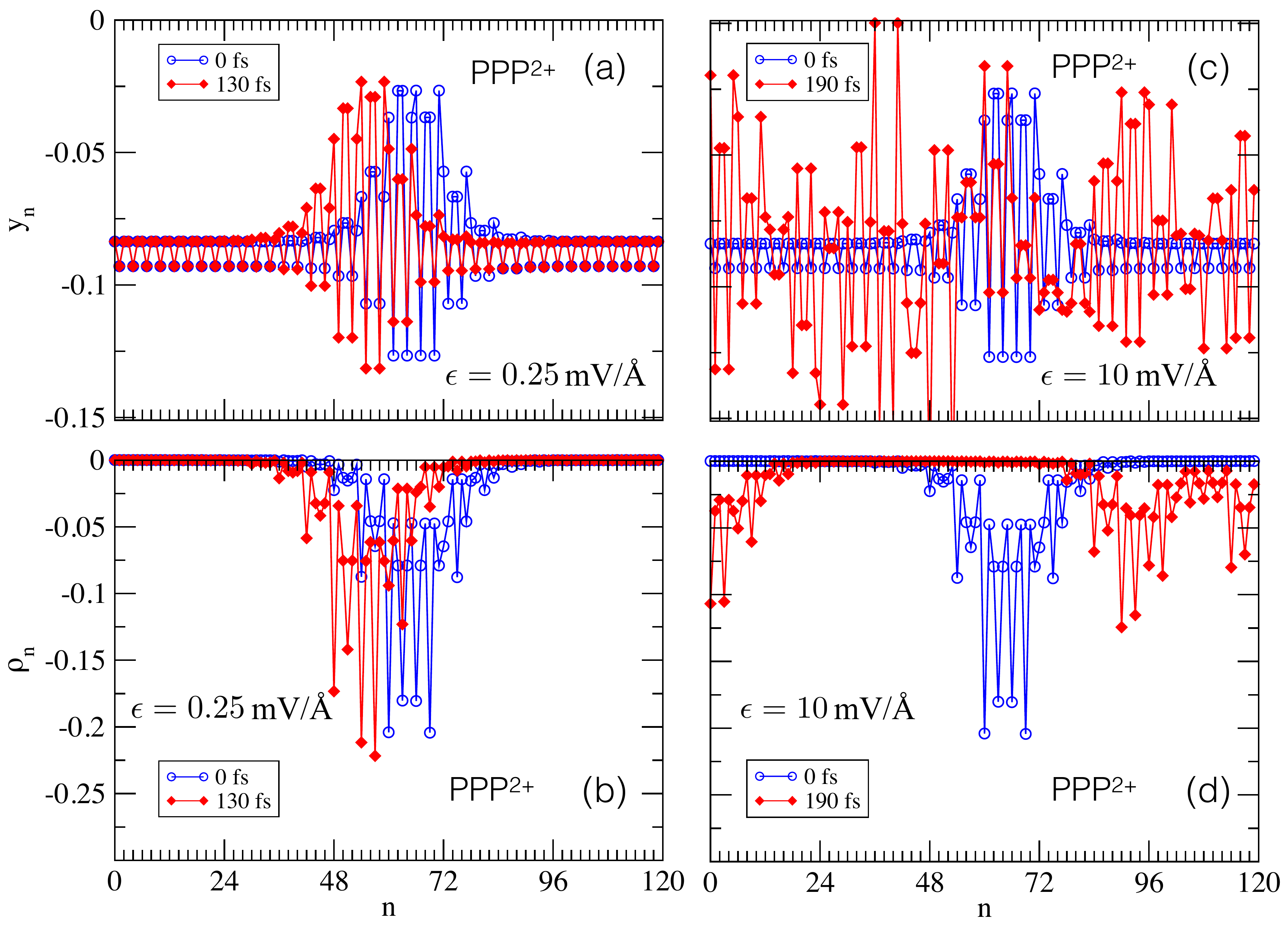}
\hspace{1.50mm}
\includegraphics[scale=0.48]{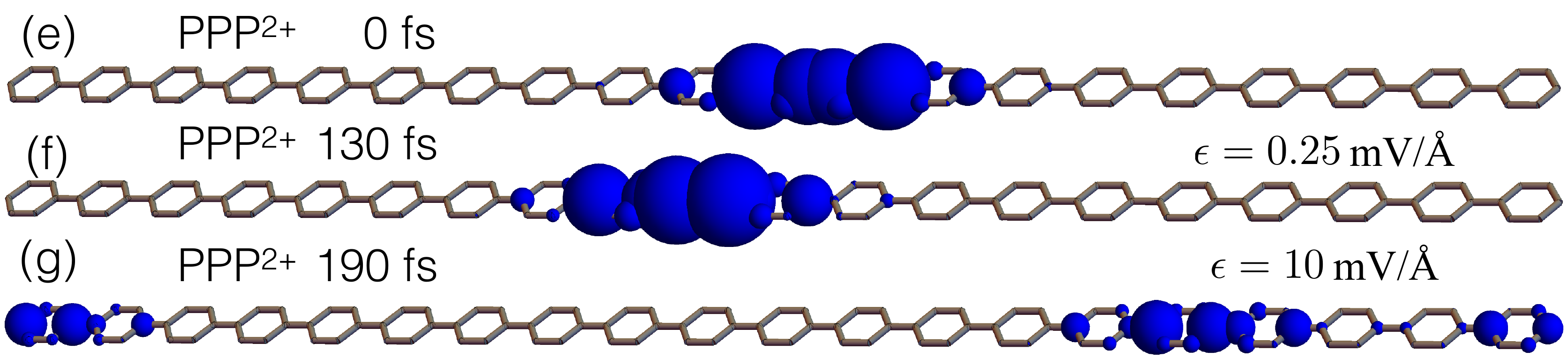}
\end{minipage}
%\vspace{0.00mm}
\caption{(Colour online) Similar to Figure~\ref{fig:wfield} but for bipolaron.
The time evolution of the lattice distortion, excess charge, and bipolaronic-state wave function of p-doped PPP after applying 0.25 and 10~mV/$\mbox{\AA}$ electric fields. (a) and (b): The evolution of the lattice distortion and the excess charge in 0.25~mV/$\mbox{\AA}$ electric field, (c) and (d): The evolution of the lattice distortion and the excess charge in 10~mV/$\mbox{\AA}$ electric field. (e)-(f) The evolution of the bipolaron wave function in 0.25~mV/$\mbox{\AA}$ electric field for two different times. (g) The snapshot of the bipolaron wave function in 10~mV/$\mbox{\AA}$ electric field. The blue line with circle is for $t=0$ and the red line with filled-diamond is for $t$. The size of the blue spheres in panels e-h is according to Fig.~\ref{fig:pol.prop}. The electric field is switched on at $t=0$ and gradually ramped up during the first 50~fs of the simulation.}
\label{fig:sfield}
\end{figure*}

We finally also briefly discuss our results for bipolarons. 
Studies for PA suggest that the exists a range of on-site Hubbard parameters~\cite{li2010electric} for which the bipolaronic state is more stable than two separate polarons. Therefore the bipolaronic state warrants consideration. In the present simulations, we neglect electron-electron correlations, as mentioned in the method section. Therefore, our bipolaron results may be considered as the limit of maximum coupling.

Fig.~\ref{fig:sfield} shows the bipolaron dynamics for selected field strengths. 
In panels a and b, the lattice displacements and excess charge for PPP$^{2+}$ in a weak field are shown. Compared to the polaron case, the displacements around the charge are localised on fewer sites and are greater in magnitude.

Figs.~\ref{fig:sfield}c-d illustrate bipolaron breakdown in a strong electric field (10~mV/$\mbox{\AA}$). Here, the bipolaron dissociates after about 190~fs. Compared to the polaron case, a significantly stronger electric field is needed to decouple the charge from the lattice distortions. However, if we were to include electron-electron correlations in the model, the bipolaronic state would become less stable, and the break-down field would in any case be lower than in the present calculation. 

\begin{figure}
\centering
\includegraphics[scale=0.33]{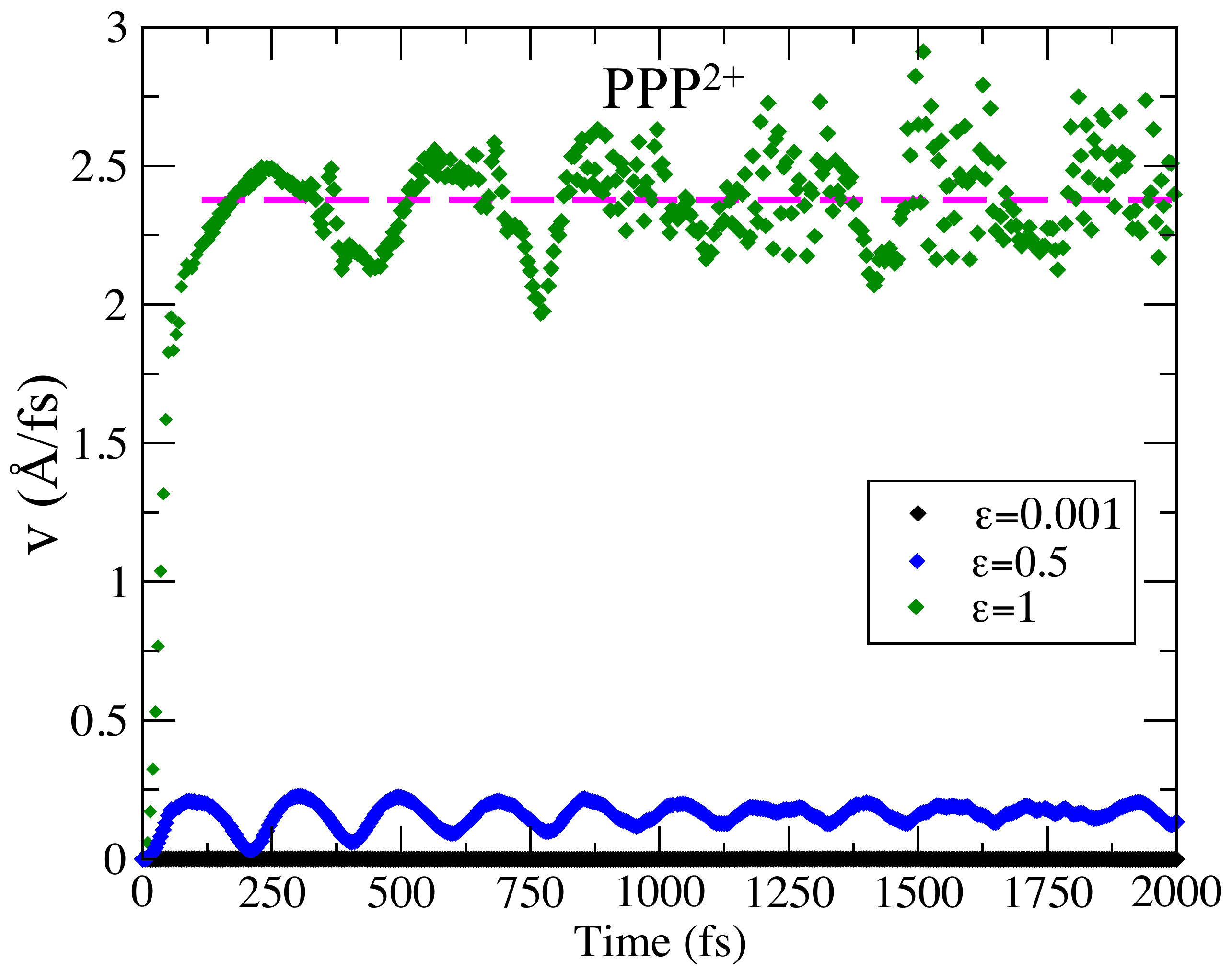}
\caption{(Colour online) The bipolaron velocity as a function of time for for three representative electric field strengths $\epsilon$ in units of~mV/$\mbox{\AA}$. The magenta dashed line is a guide for the eye.}
\label{fig:bipolvol}
\end{figure}
The bipolaron velocity for three representative field strengths is shown in Fig.~\ref{fig:bipolvol}. Just as for the polaron, the bipolaron velocity oscillates in the sonic regime (blue dots). Some initial oscillations are also discernible in the supersonic regime (green dots). The bipolaron velocity in the supersonic regime is about three times higher than for polarons. 

%% file: Conclusions.tex
\section{Conclusions}
In conclusion, we have demonstrated that polarons in fact appear to be relatively unstable in an electric field. For normal device field strengths, we find that they quickly dissolve and delocalize over the polymer. For electric field strengths just above the break-down field 1.6~mV/{\AA}, the polarons occasionally localize and then dissolve again. The results are obtained by simulation of the SSH Hamiltonian for PPP, which is an archetypal conducting polymer. Also the bipolarons in our model break down at relatively moderate electric field strengths.
The reliability of our model is ensured by comparing the PPP electronic structure for both the neutral and charged cases with available data. 
Our results challenge the common view that polarons are central for the charge transport in many types of devices based on conducting polymers.
Polaronic states can be detected using a variety of techniques, e.g., time- and wavelength-resolved pump-probe measurements, Raman spectroelectrochemistry, single-molecule fluorescence spectroscopy (SMS), photoinduced absorption (PIA) spectroscopy, and electron spin resonance (ESR) spectroscopy. We hope that the results presented here will inspire additional such experiments, explicitly addressing how the signal is affected by an external electric field.

%% file: Acknowledgment.tex
\section{acknowledgement}
%\newline
%
We acknowledge financial support from Vetenskapsr\aa det (VR), The Royal Swedish Academy of Sciences (KVA), the Knut and Alice Wallenberg Foundation (KAW), Carl Tryggers Stiftelse (CTS), Swedish Energy Agency (STEM), and Swedish Foundation for Strategic Research (SSF).
The computations were performed on resources provided by the Swedish National Infrastructure for Computing (SNIC) at the National Supercomputer Center (NSC), Link\"oping University, the PDC Centre for High Performance Computing (PDC-HPC), KTH, and the High Performance Computing Center North  (HPC2N), Ume\aa \,University.
%
%\end{acknowledgement}
%

%% file: main.bbl
\providecommand{\latin}[1]{#1}
\makeatletter
\providecommand{\doi}
  {\begingroup\let\do\@makeother\dospecials
  \catcode`\{=1 \catcode`\}=2\doi@aux}
\providecommand{\doi@aux}[1]{\endgroup\texttt{#1}}
\makeatother
\providecommand*\mcitethebibliography{\thebibliography}
\csname @ifundefined\endcsname{endmcitethebibliography}
  {\let\endmcitethebibliography\endthebibliography}{}
\begin{mcitethebibliography}{51}
\providecommand*\natexlab[1]{#1}
\providecommand*\mciteSetBstSublistMode[1]{}
\providecommand*\mciteSetBstMaxWidthForm[2]{}
\providecommand*\mciteBstWouldAddEndPuncttrue
  {\def\EndOfBibitem{\unskip.}}
\providecommand*\mciteBstWouldAddEndPunctfalse
  {\let\EndOfBibitem\relax}
\providecommand*\mciteSetBstMidEndSepPunct[3]{}
\providecommand*\mciteSetBstSublistLabelBeginEnd[3]{}
\providecommand*\EndOfBibitem{}
\mciteSetBstSublistMode{f}
\mciteSetBstMaxWidthForm{subitem}{(\alph{mcitesubitemcount})}
\mciteSetBstSublistLabelBeginEnd
  {\mcitemaxwidthsubitemform\space}
  {\relax}
  {\relax}

\bibitem[Gr{\"a}tzel(2001)]{Gratzel2001}
Gr{\"a}tzel,~M. Photoelectrochemical Cells. \emph{Nature} \textbf{2001},
  \emph{414}, 338--344\relax
\mciteBstWouldAddEndPuncttrue
\mciteSetBstMidEndSepPunct{\mcitedefaultmidpunct}
{\mcitedefaultendpunct}{\mcitedefaultseppunct}\relax
\EndOfBibitem
\bibitem[Forrest(2004)]{Forrest2004}
Forrest,~S.~R. The Path to Ubiquitous and Low-Cost Organic Electronic
  Appliances on Plastic. \emph{Nature} \textbf{2004}, \emph{428},
  911--918\relax
\mciteBstWouldAddEndPuncttrue
\mciteSetBstMidEndSepPunct{\mcitedefaultmidpunct}
{\mcitedefaultendpunct}{\mcitedefaultseppunct}\relax
\EndOfBibitem
\bibitem[Heeger \latin{et~al.}(1988)Heeger, Kivelson, Schrieffer, and
  Su]{Heeger1988}
Heeger,~A.~J.; Kivelson,~S.; Schrieffer,~J.; Su,~W.-P. Solitons in Conducting
  Polymers. \emph{Reviews of Modern Physics} \textbf{1988}, \emph{60},
  781\relax
\mciteBstWouldAddEndPuncttrue
\mciteSetBstMidEndSepPunct{\mcitedefaultmidpunct}
{\mcitedefaultendpunct}{\mcitedefaultseppunct}\relax
\EndOfBibitem
\bibitem[Gustafsson \latin{et~al.}(1992)Gustafsson, Cao, Treacy, Klavetter,
  Colaneri, and Heeger]{Gustafsson1992}
Gustafsson,~G.; Cao,~Y.; Treacy,~G.; Klavetter,~F.; Colaneri,~N.; Heeger,~A.
  Flexible Light-Emitting Diodes Made From Soluble Conducting Polymers.
  \emph{Nature} \textbf{1992}, \emph{357}, 477--479\relax
\mciteBstWouldAddEndPuncttrue
\mciteSetBstMidEndSepPunct{\mcitedefaultmidpunct}
{\mcitedefaultendpunct}{\mcitedefaultseppunct}\relax
\EndOfBibitem
\bibitem[Balint \latin{et~al.}(2014)Balint, Cassidy, and Cartmell]{Balint2014}
Balint,~R.; Cassidy,~N.~J.; Cartmell,~S.~H. Conductive Polymers: Towards a
  Smart Biomaterial for Tissue Engineering. \emph{Acta biomaterialia}
  \textbf{2014}, \emph{10}, 2341--2353\relax
\mciteBstWouldAddEndPuncttrue
\mciteSetBstMidEndSepPunct{\mcitedefaultmidpunct}
{\mcitedefaultendpunct}{\mcitedefaultseppunct}\relax
\EndOfBibitem
\bibitem[Vasseur \latin{et~al.}(2016)Vasseur, Fagot-Revurat, Sicot, Kierren,
  Moreau, Malterre, Cardenas, Galeotti, Lipton-Duffin, Rosei, \latin{et~al.}
  others]{Vasseur2016}
Vasseur,~G.; Fagot-Revurat,~Y.; Sicot,~M.; Kierren,~B.; Moreau,~L.;
  Malterre,~D.; Cardenas,~L.; Galeotti,~G.; Lipton-Duffin,~J.; Rosei,~F.
  \latin{et~al.}  Quasi One-Dimensional Band Dispersion and Surface
  Metallization in Long-Range Ordered Polymeric Wires. \emph{Nature
  communications} \textbf{2016}, \emph{7}\relax
\mciteBstWouldAddEndPuncttrue
\mciteSetBstMidEndSepPunct{\mcitedefaultmidpunct}
{\mcitedefaultendpunct}{\mcitedefaultseppunct}\relax
\EndOfBibitem
\bibitem[Pandey \latin{et~al.}(2015)Pandey, Yadav, Upadhyay, Prakash, and
  Mishra]{pandey2015surface}
Pandey,~R.~K.; Yadav,~S.~K.; Upadhyay,~C.; Prakash,~R.; Mishra,~H. Surface
  Plasmon Coupled Metal Enhanced Spectral and Charge Transport Properties of
  Poly(3,3-dialkylquarterthiophene) Langmuir Schaefer Films. \emph{Nanoscale}
  \textbf{2015}, \emph{7}, 6083--6092\relax
\mciteBstWouldAddEndPuncttrue
\mciteSetBstMidEndSepPunct{\mcitedefaultmidpunct}
{\mcitedefaultendpunct}{\mcitedefaultseppunct}\relax
\EndOfBibitem
\bibitem[Chen \latin{et~al.}(2014)Chen, Shao, Xiao, Rondinone, Loo, Kent,
  Sumpter, Li, Keum, Diemer, \latin{et~al.} others]{chen2014solvent}
Chen,~J.; Shao,~M.; Xiao,~K.; Rondinone,~A.~J.; Loo,~Y.-L.; Kent,~P.~R.;
  Sumpter,~B.~G.; Li,~D.; Keum,~J.~K.; Diemer,~P.~J. \latin{et~al.}
  Solvent-Type-Dependent Polymorphism and Charge Transport in a Long Fused-Ring
  Organic Semiconductor. \emph{Nanoscale} \textbf{2014}, \emph{6},
  449--456\relax
\mciteBstWouldAddEndPuncttrue
\mciteSetBstMidEndSepPunct{\mcitedefaultmidpunct}
{\mcitedefaultendpunct}{\mcitedefaultseppunct}\relax
\EndOfBibitem
\bibitem[Adhikary \latin{et~al.}(2013)Adhikary, Venkatesan, Adhikari, Maharjan,
  Adebanjo, Chen, and Qiao]{adhikary2013enhanced}
Adhikary,~P.; Venkatesan,~S.; Adhikari,~N.; Maharjan,~P.~P.; Adebanjo,~O.;
  Chen,~J.; Qiao,~Q. Enhanced Charge Transport and Photovoltaic Performance of
  PBDTTT-CT/PC 70 BM Solar Cells via UV--Ozone Treatment. \emph{Nanoscale}
  \textbf{2013}, \emph{5}, 10007--10013\relax
\mciteBstWouldAddEndPuncttrue
\mciteSetBstMidEndSepPunct{\mcitedefaultmidpunct}
{\mcitedefaultendpunct}{\mcitedefaultseppunct}\relax
\EndOfBibitem
\bibitem[Heimel(2016)]{heimel2016optical}
Heimel,~G. The Optical Signature of Charges in Conjugated Polymers. \emph{ACS
  central science} \textbf{2016}, \relax
\mciteBstWouldAddEndPunctfalse
\mciteSetBstMidEndSepPunct{\mcitedefaultmidpunct}
{}{\mcitedefaultseppunct}\relax
\EndOfBibitem
\bibitem[Junior and Stafstr{\"o}m(2015)Junior, and Stafstr{\"o}m]{Ribeiro2015}
Junior,~L. A.~R.; Stafstr{\"o}m,~S. Polaron Stability in Molecular
  Semiconductors: Theoretical Insight into the Impact of the Temperature,
  Electric Field and the System Dimensionality. \emph{Physical Chemistry
  Chemical Physics} \textbf{2015}, \emph{17}, 8973--8982\relax
\mciteBstWouldAddEndPuncttrue
\mciteSetBstMidEndSepPunct{\mcitedefaultmidpunct}
{\mcitedefaultendpunct}{\mcitedefaultseppunct}\relax
\EndOfBibitem
\bibitem[Ribeiro \latin{et~al.}(2013)Ribeiro, da~Cunha, de~Oliveria~Neto,
  Gargano, and e~Silva]{Ribeiro2013}
Ribeiro,~L.~A.; da~Cunha,~W.~F.; de~Oliveria~Neto,~P.~H.; Gargano,~R.;
  e~Silva,~G.~M. Effects of Temperature and Electric Field Induced Phase
  Transitions on the Dynamics of Polarons and Bipolarons. \emph{New Journal of
  Chemistry} \textbf{2013}, \emph{37}, 2829--2836\relax
\mciteBstWouldAddEndPuncttrue
\mciteSetBstMidEndSepPunct{\mcitedefaultmidpunct}
{\mcitedefaultendpunct}{\mcitedefaultseppunct}\relax
\EndOfBibitem
\bibitem[da~Silva \latin{et~al.}(2012)da~Silva, de~Oliveira~Neto, da~Cunha,
  Gargano, and e~Silva]{Silva2012}
da~Silva,~M. V.~A.; de~Oliveira~Neto,~P.~H.; da~Cunha,~W.~F.; Gargano,~R.;
  e~Silva,~G.~M. Supersonic Quasi-Particles Dynamics in Organic Semiconductors.
  \emph{Chemical Physics Letters} \textbf{2012}, \emph{550}, 146--149\relax
\mciteBstWouldAddEndPuncttrue
\mciteSetBstMidEndSepPunct{\mcitedefaultmidpunct}
{\mcitedefaultendpunct}{\mcitedefaultseppunct}\relax
\EndOfBibitem
\bibitem[Johansson and Stafstr{\"o}m(2004)Johansson, and
  Stafstr{\"o}m]{Johansson2004}
Johansson,~A.~A.; Stafstr{\"o}m,~S. Nonadiabatic Simulations of Polaron
  Dynamics. \emph{Physical Review B} \textbf{2004}, \emph{69}, 235205\relax
\mciteBstWouldAddEndPuncttrue
\mciteSetBstMidEndSepPunct{\mcitedefaultmidpunct}
{\mcitedefaultendpunct}{\mcitedefaultseppunct}\relax
\EndOfBibitem
\bibitem[Zhang(2002)]{zhang2002microscopic}
Zhang,~S. The Microscopic Origin of the Doping Limits in Semiconductors and
  Wide-Gap Materials and Recent Developments in Overcoming these Limits: a
  Review. \emph{Journal of Physics: Condensed Matter} \textbf{2002}, \emph{14},
  R881\relax
\mciteBstWouldAddEndPuncttrue
\mciteSetBstMidEndSepPunct{\mcitedefaultmidpunct}
{\mcitedefaultendpunct}{\mcitedefaultseppunct}\relax
\EndOfBibitem
\bibitem[Ivory \latin{et~al.}(1979)Ivory, Miller, Sowa, Shacklette, Chance, and
  Baughman]{ivory1979highly}
Ivory,~D.; Miller,~G.; Sowa,~J.; Shacklette,~L.; Chance,~R.; Baughman,~R.
  Highly Conducting Charge-Transfer Complexes of Poly(p-phenylene). \emph{The
  Journal of Chemical Physics} \textbf{1979}, \emph{71}, 1506--1507\relax
\mciteBstWouldAddEndPuncttrue
\mciteSetBstMidEndSepPunct{\mcitedefaultmidpunct}
{\mcitedefaultendpunct}{\mcitedefaultseppunct}\relax
\EndOfBibitem
\bibitem[Kovacic and Jones(1987)Kovacic, and Jones]{kovacic1987dehydro}
Kovacic,~P.; Jones,~M.~B. Dehydro Coupling of Aromatic Nuclei by
  Catalyst-Oxidant Systems: Poly(p-phenylene). \emph{Chemical Reviews}
  \textbf{1987}, \emph{87}, 357--379\relax
\mciteBstWouldAddEndPuncttrue
\mciteSetBstMidEndSepPunct{\mcitedefaultmidpunct}
{\mcitedefaultendpunct}{\mcitedefaultseppunct}\relax
\EndOfBibitem
\bibitem[Grem \latin{et~al.}(1992)Grem, Leditzky, Ullrich, and
  Leising]{grem1992realization}
Grem,~G.; Leditzky,~G.; Ullrich,~B.; Leising,~G. Realization of a
  Blue-Light-Emitting Device Using Poly(p-phenylene). \emph{Advanced Materials}
  \textbf{1992}, \emph{4}, 36--37\relax
\mciteBstWouldAddEndPuncttrue
\mciteSetBstMidEndSepPunct{\mcitedefaultmidpunct}
{\mcitedefaultendpunct}{\mcitedefaultseppunct}\relax
\EndOfBibitem
\bibitem[Grem \latin{et~al.}(1992)Grem, Leditzky, Ullrich, and
  Leising]{grem1992blue}
Grem,~G.; Leditzky,~G.; Ullrich,~B.; Leising,~G. Blue Electroluminescent Device
  Based on a Conjugated Polymer. \emph{Synthetic metals} \textbf{1992},
  \emph{51}, 383--389\relax
\mciteBstWouldAddEndPuncttrue
\mciteSetBstMidEndSepPunct{\mcitedefaultmidpunct}
{\mcitedefaultendpunct}{\mcitedefaultseppunct}\relax
\EndOfBibitem
\bibitem[Grem and Leising(1993)Grem, and Leising]{grem1993electroluminescence}
Grem,~G.; Leising,~G. Electroluminescence of “Wide-Bandgap” Chemically
  Tunable Cyclic Conjugated Polymers. \emph{Synthetic metals} \textbf{1993},
  \emph{57}, 4105--4110\relax
\mciteBstWouldAddEndPuncttrue
\mciteSetBstMidEndSepPunct{\mcitedefaultmidpunct}
{\mcitedefaultendpunct}{\mcitedefaultseppunct}\relax
\EndOfBibitem
\bibitem[Johansson and Stafstr{\"o}m(2001)Johansson, and
  Stafstr{\"o}m]{Johansson2001}
Johansson,~{\AA}.; Stafstr{\"o}m,~S. Polaron Dynamics in a System of Coupled
  Conjugated Polymer Chains. \emph{Physical review letters} \textbf{2001},
  \emph{86}, 3602\relax
\mciteBstWouldAddEndPuncttrue
\mciteSetBstMidEndSepPunct{\mcitedefaultmidpunct}
{\mcitedefaultendpunct}{\mcitedefaultseppunct}\relax
\EndOfBibitem
\bibitem[Basko and Conwell(2002)Basko, and Conwell]{Basko2002}
Basko,~D.; Conwell,~E. Stationary Polaron Motion in a Polymer Chain at High
  Electric Fields. \emph{Physical review letters} \textbf{2002}, \emph{88},
  056401\relax
\mciteBstWouldAddEndPuncttrue
\mciteSetBstMidEndSepPunct{\mcitedefaultmidpunct}
{\mcitedefaultendpunct}{\mcitedefaultseppunct}\relax
\EndOfBibitem
\bibitem[Su \latin{et~al.}(1979)Su, Schrieffer, and Heeger]{su1979solitons}
Su,~W.; Schrieffer,~J.; Heeger,~A.~J. Solitons in Polyacetylene. \emph{Physical
  Review Letters} \textbf{1979}, \emph{42}, 1698\relax
\mciteBstWouldAddEndPuncttrue
\mciteSetBstMidEndSepPunct{\mcitedefaultmidpunct}
{\mcitedefaultendpunct}{\mcitedefaultseppunct}\relax
\EndOfBibitem
\bibitem[Su and Schrieffer(1981)Su, and Schrieffer]{su1981fractionally}
Su,~W.; Schrieffer,~J. Fractionally Charged Excitations in Charge-Density-Wave
  Systems with Commensurability 3. \emph{Physical Review Letters}
  \textbf{1981}, \emph{46}, 738\relax
\mciteBstWouldAddEndPuncttrue
\mciteSetBstMidEndSepPunct{\mcitedefaultmidpunct}
{\mcitedefaultendpunct}{\mcitedefaultseppunct}\relax
\EndOfBibitem
\bibitem[Su \latin{et~al.}(1980)Su, Schrieffer, and Heeger]{PhysRevB.22.2099}
Su,~W.~P.; Schrieffer,~J.~R.; Heeger,~A.~J. Soliton Excitations in
  Polyacetylene. \emph{Physical Review B} \textbf{1980}, \emph{22},
  2099--2111\relax
\mciteBstWouldAddEndPuncttrue
\mciteSetBstMidEndSepPunct{\mcitedefaultmidpunct}
{\mcitedefaultendpunct}{\mcitedefaultseppunct}\relax
\EndOfBibitem
\bibitem[Takayama \latin{et~al.}(1980)Takayama, Lin-Liu, and
  Maki]{PhysRevB.21.2388}
Takayama,~H.; Lin-Liu,~Y.~R.; Maki,~K. Continuum Model for Solitons in
  Polyacetylene. \emph{Physical Review B} \textbf{1980}, \emph{21},
  2388--2393\relax
\mciteBstWouldAddEndPuncttrue
\mciteSetBstMidEndSepPunct{\mcitedefaultmidpunct}
{\mcitedefaultendpunct}{\mcitedefaultseppunct}\relax
\EndOfBibitem
\bibitem[Su and Schrieffer(1980)Su, and Schrieffer]{su1980soliton}
Su,~W.; Schrieffer,~J. Soliton Dynamics in Polyacetylene. \emph{Proceedings of
  the National Academy of Sciences} \textbf{1980}, \emph{77}, 5626--5629\relax
\mciteBstWouldAddEndPuncttrue
\mciteSetBstMidEndSepPunct{\mcitedefaultmidpunct}
{\mcitedefaultendpunct}{\mcitedefaultseppunct}\relax
\EndOfBibitem
\bibitem[Ono and Terai(1990)Ono, and Terai]{ono1990motion}
Ono,~Y.; Terai,~A. Motion of Charged Soliton in Polyacetylene Due to Electric
  Field. \emph{Journal of the Physical Society of Japan} \textbf{1990},
  \emph{59}, 2893--2904\relax
\mciteBstWouldAddEndPuncttrue
\mciteSetBstMidEndSepPunct{\mcitedefaultmidpunct}
{\mcitedefaultendpunct}{\mcitedefaultseppunct}\relax
\EndOfBibitem
\bibitem[Bubnova \latin{et~al.}(2012)Bubnova, Berggren, and
  Crispin]{bubnova2012tuning}
Bubnova,~O.; Berggren,~M.; Crispin,~X. Tuning the Thermoelectric Properties of
  Conducting Polymers in an Electrochemical Transistor. \emph{Journal of the
  American Chemical Society} \textbf{2012}, \emph{134}, 16456--16459\relax
\mciteBstWouldAddEndPuncttrue
\mciteSetBstMidEndSepPunct{\mcitedefaultmidpunct}
{\mcitedefaultendpunct}{\mcitedefaultseppunct}\relax
\EndOfBibitem
\bibitem[Lin \latin{et~al.}(2003)Lin, Meng, Shy, Horng, Yu, Chen, Liaw, Huang,
  Peng, and Chen]{lin2003triplet}
Lin,~L.; Meng,~H.; Shy,~J.; Horng,~S.; Yu,~L.; Chen,~C.; Liaw,~H.; Huang,~C.;
  Peng,~K.; Chen,~S. Triplet-To-Singlet Exciton Formation in
  Poly(p-phenylene-vinylene) Light-Emitting Diodes. \emph{Physical review
  letters} \textbf{2003}, \emph{90}, 036601\relax
\mciteBstWouldAddEndPuncttrue
\mciteSetBstMidEndSepPunct{\mcitedefaultmidpunct}
{\mcitedefaultendpunct}{\mcitedefaultseppunct}\relax
\EndOfBibitem
\bibitem[Gross \latin{et~al.}(2000)Gross, M{\"u}ller, Nothofer, Scherf, Neher,
  Br{\"a}uchle, and Meerholz]{gross2000improving}
Gross,~M.; M{\"u}ller,~D.~C.; Nothofer,~H.-G.; Scherf,~U.; Neher,~D.;
  Br{\"a}uchle,~C.; Meerholz,~K. Improving the Performance of Doped
  $\pi$-Conjugated Polymers for Use in Organic Light-Emitting Diodes.
  \emph{Nature} \textbf{2000}, \emph{405}, 661--665\relax
\mciteBstWouldAddEndPuncttrue
\mciteSetBstMidEndSepPunct{\mcitedefaultmidpunct}
{\mcitedefaultendpunct}{\mcitedefaultseppunct}\relax
\EndOfBibitem
\bibitem[Barth \latin{et~al.}(2001)Barth, M{\"u}ller, Riel, Seidler, Riess,
  Vestweber, and B{\"a}ssler]{barth2001electron}
Barth,~S.; M{\"u}ller,~P.; Riel,~H.; Seidler,~P.; Riess,~W.; Vestweber,~H.;
  B{\"a}ssler,~H. Electron Mobility in Tris(8-hydroxy-quinoline) Aluminum Thin
  Films Determined via Transient Electroluminescence from Single- and
  Multilayer Organic Light-Emitting Diodes. \emph{Journal of Applied physics}
  \textbf{2001}, \emph{89}, 3711--3719\relax
\mciteBstWouldAddEndPuncttrue
\mciteSetBstMidEndSepPunct{\mcitedefaultmidpunct}
{\mcitedefaultendpunct}{\mcitedefaultseppunct}\relax
\EndOfBibitem
\bibitem[Yamamoto \latin{et~al.}(2005)Yamamoto, Wilkinson, Long, Bussman,
  Christodoulides, and Kafafi]{yamamoto2005nanoscale}
Yamamoto,~H.; Wilkinson,~J.; Long,~J.~P.; Bussman,~K.; Christodoulides,~J.~A.;
  Kafafi,~Z.~H. Nanoscale Organic Light-Emitting Diodes. \emph{Nano letters}
  \textbf{2005}, \emph{5}, 2485--2488\relax
\mciteBstWouldAddEndPuncttrue
\mciteSetBstMidEndSepPunct{\mcitedefaultmidpunct}
{\mcitedefaultendpunct}{\mcitedefaultseppunct}\relax
\EndOfBibitem
\bibitem[Tuti{\^s} \latin{et~al.}(2003)Tuti{\^s}, Berner, and
  Zuppiroli]{tutis2003internal}
Tuti{\^s},~E.; Berner,~D.; Zuppiroli,~L. Internal Electric Field and Charge
  Distribution in Multilayer Organic Light-Emitting Diodes. \emph{Journal of
  applied physics} \textbf{2003}, \emph{93}, 4594--4602\relax
\mciteBstWouldAddEndPuncttrue
\mciteSetBstMidEndSepPunct{\mcitedefaultmidpunct}
{\mcitedefaultendpunct}{\mcitedefaultseppunct}\relax
\EndOfBibitem
\bibitem[com()]{comment1}
Of course, the carbon atoms linking together adjacent phenyl rings are not
  CH-groups, but for simplicity of the model this mass is assumed for all
  sites.\relax
\mciteBstWouldAddEndPunctfalse
\mciteSetBstMidEndSepPunct{\mcitedefaultmidpunct}
{}{\mcitedefaultseppunct}\relax
\EndOfBibitem
\bibitem[Peierls(1955)]{peierls1955quantum}
Peierls,~R.~E. \emph{Quantum Theory of Solids}; Oxford University Press,
  1955\relax
\mciteBstWouldAddEndPuncttrue
\mciteSetBstMidEndSepPunct{\mcitedefaultmidpunct}
{\mcitedefaultendpunct}{\mcitedefaultseppunct}\relax
\EndOfBibitem
\bibitem[Johansson and Stafstr{\"o}m(2004)Johansson, and
  Stafstr{\"o}m]{johansson2004nonadiabatic}
Johansson,~A.~A.; Stafstr{\"o}m,~S. Nonadiabatic Simulations of Polaron
  Dynamics. \emph{Physical Review B} \textbf{2004}, \emph{69}, 235205\relax
\mciteBstWouldAddEndPuncttrue
\mciteSetBstMidEndSepPunct{\mcitedefaultmidpunct}
{\mcitedefaultendpunct}{\mcitedefaultseppunct}\relax
\EndOfBibitem
\bibitem[Diaz \latin{et~al.}(1981)Diaz, Crowley, Bargon, Gardini, and
  Torrance]{diaz1981}
Diaz,~A.; Crowley,~J.; Bargon,~J.; Gardini,~G.; Torrance,~J. Electrooxidation
  of Aromatic Oligomers and Conducting Polymers. \emph{Journal of
  Electroanalytical Chemistry and Interfacial Electrochemistry} \textbf{1981},
  \emph{121}, 355--361\relax
\mciteBstWouldAddEndPuncttrue
\mciteSetBstMidEndSepPunct{\mcitedefaultmidpunct}
{\mcitedefaultendpunct}{\mcitedefaultseppunct}\relax
\EndOfBibitem
\bibitem[Botelho \latin{et~al.}(2011)Botelho, Shin, Li, Jiang, and
  Lin]{botelho2011unified}
Botelho,~A.~L.; Shin,~Y.; Li,~M.; Jiang,~L.; Lin,~X. Unified Hamiltonian for
  Conducting Polymers. \emph{Journal of Physics: Condensed Matter}
  \textbf{2011}, \emph{23}, 455501\relax
\mciteBstWouldAddEndPuncttrue
\mciteSetBstMidEndSepPunct{\mcitedefaultmidpunct}
{\mcitedefaultendpunct}{\mcitedefaultseppunct}\relax
\EndOfBibitem
\bibitem[Rossi(1991)]{rossi1991minimum}
Rossi,~G. Minimum Total Energy Calculations for Conjugated Polymer Chains.
  \emph{The Journal of chemical physics} \textbf{1991}, \emph{94},
  4031--4041\relax
\mciteBstWouldAddEndPuncttrue
\mciteSetBstMidEndSepPunct{\mcitedefaultmidpunct}
{\mcitedefaultendpunct}{\mcitedefaultseppunct}\relax
\EndOfBibitem
\bibitem[Kuwabara \latin{et~al.}(1991)Kuwabara, Ono, and
  Terai]{kuwabara1991motion}
Kuwabara,~M.; Ono,~Y.; Terai,~A. Motion of Charged Soliton in Polyacetylene Due
  to Electric Field. II. Behavior of Width. \emph{Journal of the Physical
  Society of Japan} \textbf{1991}, \emph{60}, 1286--1293\relax
\mciteBstWouldAddEndPuncttrue
\mciteSetBstMidEndSepPunct{\mcitedefaultmidpunct}
{\mcitedefaultendpunct}{\mcitedefaultseppunct}\relax
\EndOfBibitem
\bibitem[Rakhmanova and Conwell(1999)Rakhmanova, and
  Conwell]{rakhmanova1999polaron}
Rakhmanova,~S.; Conwell,~E. Polaron Dissociation in Conducting Polymers by High
  Electric Fields. \emph{Applied physics letters} \textbf{1999}, \emph{75},
  1518--1520\relax
\mciteBstWouldAddEndPuncttrue
\mciteSetBstMidEndSepPunct{\mcitedefaultmidpunct}
{\mcitedefaultendpunct}{\mcitedefaultseppunct}\relax
\EndOfBibitem
\bibitem[Ozawa and Ono(2000)Ozawa, and Ono]{ozawa2000dynamics}
Ozawa,~T.; Ono,~Y. Dynamics of an Acoustic Bipolaron in One-Dimensional
  Electron-Lattice Systems. \emph{Journal of the Physical Society of Japan}
  \textbf{2000}, \emph{69}, 1162--1169\relax
\mciteBstWouldAddEndPuncttrue
\mciteSetBstMidEndSepPunct{\mcitedefaultmidpunct}
{\mcitedefaultendpunct}{\mcitedefaultseppunct}\relax
\EndOfBibitem
\bibitem[Subbaswamy and Grabowski(1981)Subbaswamy, and
  Grabowski]{PhysRevB.24.2168}
Subbaswamy,~K.~R.; Grabowski,~M. Bond Alternation, On-site Coulomb
  Correlations, and Solitons in Polyacetylene. \emph{Physical Review B}
  \textbf{1981}, \emph{24}, 2168--2173\relax
\mciteBstWouldAddEndPuncttrue
\mciteSetBstMidEndSepPunct{\mcitedefaultmidpunct}
{\mcitedefaultendpunct}{\mcitedefaultseppunct}\relax
\EndOfBibitem
\bibitem[Lin \latin{et~al.}(2006)Lin, Li, F{\"o}rst, and Yip]{lin2006multiple}
Lin,~X.; Li,~J.; F{\"o}rst,~C.~J.; Yip,~S. Multiple Self-Localized Electronic
  States in Trans-Polyacetylene. \emph{Proceedings of the National Academy of
  Sciences} \textbf{2006}, \emph{103}, 8943--8946\relax
\mciteBstWouldAddEndPuncttrue
\mciteSetBstMidEndSepPunct{\mcitedefaultmidpunct}
{\mcitedefaultendpunct}{\mcitedefaultseppunct}\relax
\EndOfBibitem
\bibitem[Jeckelmann and White(1998)Jeckelmann, and
  White]{jeckelmann1998density}
Jeckelmann,~E.; White,~S.~R. Density-Matrix Renormalization-Group Study of the
  Polaron problem in the Holstein Model. \emph{Physical Review B}
  \textbf{1998}, \emph{57}, 6376\relax
\mciteBstWouldAddEndPuncttrue
\mciteSetBstMidEndSepPunct{\mcitedefaultmidpunct}
{\mcitedefaultendpunct}{\mcitedefaultseppunct}\relax
\EndOfBibitem
\bibitem[Puschnig \latin{et~al.}(2012)Puschnig, Amiri, and
  Draxl]{puschnig2012band}
Puschnig,~P.; Amiri,~P.; Draxl,~C. Band Renormalization of a Polymer
  Physisorbed on Graphene Investigated by Many-Body Perturbation Theory.
  \emph{Physical Review B} \textbf{2012}, \emph{86}, 085107\relax
\mciteBstWouldAddEndPuncttrue
\mciteSetBstMidEndSepPunct{\mcitedefaultmidpunct}
{\mcitedefaultendpunct}{\mcitedefaultseppunct}\relax
\EndOfBibitem
\bibitem[Eckhardt \latin{et~al.}(1989)Eckhardt, Shacklette, Jen, and
  Elsenbaumer]{eckhardt1989electronic}
Eckhardt,~H.; Shacklette,~L.; Jen,~K.; Elsenbaumer,~R. The Electronic and
  Electrochemical Properties of Poly(phenylene vinylenes) and Poly(thienylene
  vinylenes): An Experimental and Theoretical Study. \emph{The Journal of
  Chemical Physics} \textbf{1989}, \emph{91}, 1303--1315\relax
\mciteBstWouldAddEndPuncttrue
\mciteSetBstMidEndSepPunct{\mcitedefaultmidpunct}
{\mcitedefaultendpunct}{\mcitedefaultseppunct}\relax
\EndOfBibitem
\bibitem[Liu \latin{et~al.}(2007)Liu, Gao, Li, Fu, Wei, and
  Xie]{liu2007polaron}
Liu,~X.; Gao,~K.; Li,~Y.; Fu,~J.; Wei,~J.; Xie,~S. Polaron Formation Dynamics
  in Conducting Polymers. \emph{Synthetic metals} \textbf{2007}, \emph{157},
  380--385\relax
\mciteBstWouldAddEndPuncttrue
\mciteSetBstMidEndSepPunct{\mcitedefaultmidpunct}
{\mcitedefaultendpunct}{\mcitedefaultseppunct}\relax
\EndOfBibitem
\bibitem[Li and Lagowski(2010)Li, and Lagowski]{li2010electric}
Li,~Y.; Lagowski,~J.~B. Electric Field Effects on Bipolaron Transport in
  Heterocyclic Conjugated Polymers with Application to Polythiophene.
  \emph{Optical Materials} \textbf{2010}, \emph{32}, 1177--1187\relax
\mciteBstWouldAddEndPuncttrue
\mciteSetBstMidEndSepPunct{\mcitedefaultmidpunct}
{\mcitedefaultendpunct}{\mcitedefaultseppunct}\relax
\EndOfBibitem
\end{mcitethebibliography}
